\documentclass[twocolumn]{aastex61}
\pdfoutput=1 
\usepackage{amsmath,amstext}
\usepackage[T1]{fontenc}
\usepackage{apjfonts} 
\usepackage[figure,figure*]{hypcap}
\usepackage{booktabs}
\usepackage{upgreek}
\usepackage{longtable}
\usepackage{comment}
\usepackage{upgreek}
\usepackage[caption=false]{subfig}
\usepackage{float}
\newcommand\aastex{AAS\TeX}

\newcommand{\frb}{FRB~121102}
\received{}
\revised{}
\accepted{}

\shorttitle{\frb\ low energy bursts}
\shortauthors{Gourdji et al.}

\begin{document}

\title{A sample of low energy bursts from FRB 121102}

\correspondingauthor{K.~Gourdji}
\email{k.gourdji@uva.nl}
\author{K.~Gourdji}
\affiliation{Anton Pannekoek Institute for Astronomy, University of Amsterdam, Science Park 904, 1098~XH Amsterdam, The~Netherlands}
\author{D.~Michilli}
\affiliation{Department of Physics, McGill University, 3600 rue University, Montr\'eal, QC H3A 2T8, Canada}
\affiliation{McGill Space Institute, McGill University, 3550 rue University, Montr\'eal, QC H3A 2A7, Canada}
\affiliation{Anton Pannekoek Institute for Astronomy, University of Amsterdam, Science Park 904, 1098~XH Amsterdam, The~Netherlands}
\affiliation{ASTRON, Netherlands Institute for Radio Astronomy, Oude Hoogeveensedijk 4, 7991 PD Dwingeloo, The Netherlands}
\author{L.~G.~Spitler}
\affiliation{Max-Planck-Institut f\"ur Radioastronomie, Auf dem H\"{u}gel 69, D-53121, Bonn, Germany}
\author{J.~W.~T.~Hessels}
\affiliation{Anton Pannekoek Institute for Astronomy, University of Amsterdam, Science Park 904, 1098~XH Amsterdam, The~Netherlands}
\affiliation{ASTRON, Netherlands Institute for Radio Astronomy, Oude Hoogeveensedijk 4, 7991 PD Dwingeloo, The Netherlands}
\author{A.~Seymour}
\affiliation{Green Bank Observatory, PO Box 2, Green Bank, WV 24494, USA}
\author{J.M~Cordes}
\affiliation{Cornell Center for Astrophysics and Planetary Science, Cornell University, Ithaca, NY 14853, USA}
\author{S.~Chatterjee}
\affiliation{Cornell Center for Astrophysics and Planetary Science, Cornell University, Ithaca, NY 14853, USA}




\begin{abstract}
We present 41 bursts from the first repeating fast radio burst discovered (FRB~121102). A deep search has allowed us to probe unprecedentedly low burst energies during two consecutive observations (separated by one day) using the Arecibo telescope at 1.4\,GHz. The bursts are generally detected in less than a third of the 580-MHz observing bandwidth, demonstrating that narrow-band FRB signals may be more common than previously thought. We show that the bursts are likely faint versions of previously reported multi-component bursts. There is a striking lack of bursts detected below 1.35\,GHz and simultaneous VLA observations at 3\,GHz did not detect any of the 41 bursts, but did detect one that was not seen with Arecibo, suggesting preferred radio emission frequencies that vary with epoch. A power law approximation of the cumulative distribution of burst energies yields an index $-1.8\pm0.3$ that is much steeper than the previously reported value of $\sim-0.7$. The discrepancy may be evidence for a more complex energy distribution. We place constraints on the possibility that the associated persistent radio source is generated by the emission of many faint bursts ($\sim700$\,ms$^{-1}$). We do not see a connection between burst fluence and wait time. The distribution of wait times follows a log-normal distribution centered around $\sim200$\,s; however, some bursts have wait times below 1\,s and as short as 26\,ms, which is consistent with previous reports of a bimodal distribution. We caution against exclusively integrating over the full observing band during FRB searches, because this can lower signal-to-noise.
\end{abstract}




\section{Introduction} \label{sec:intro}
Fast radio bursts (FRBs) are bright (peak flux density $0.01 - 100$\,Jy), millisecond-duration radio pulses of extragalactic origin \citep{thornton2013population}. The physical source of FRBs has been a mystery since the first example was discovered over ten years ago \citep{lorimer2007bright}. The bursts must arise from coherent radiation from a small emission region, and both cataclysmic explosions and longer-lived progenitors have been hypothesized (see the reviews by \citealt{popov18rev,katz18review,platts18} and references therein). To date, over 60 FRB sources have appeared in the literature\footnote{An overview of all published FRBs and their properties is provided in a catalogue by \cite{frbcat} available at \url{ http://www.frbcat.org}.}. All have a large dispersion measure (DM), in excess of the expected Galactic contribution along the line of site \citep{cordes2002ne2001}, which, in the absence of a host-galaxy association, is the primary evidence for their extragalactic origin.
The observed durations of the bursts range between $0.03$--$26$\,ms \citep{michilli2018,FarahAtel17} and they have been detected over a reasonably large fractional bandwidth, which can teach us about their spectra as long as instrumental effects do not dominate. 
The bandwidth of the recording systems that have detected FRBs ranges from 16\,MHz to 580\,MHz and is as large as 4\,GHz for observations of one source \citep{gajjar18,zhang18}. 

Some bursts have been detected with (sub-)millisecond temporal structure \citep{championFiveFRBs,Farah18,shannon18,hessels18} thanks to (in some cases) real-time coherent de-dispersion or raw-voltage capture. However, the observed durations of
most bursts are limited by the recorded time resolution, intra-channel dispersive smearing or scattering in some cases, making it difficult to study their spectro-temporal structure \citep{Bhandari18}.

Despite extensive follow-up observations (e.g. \citealt{petroff2015repeatability,shannon18}), only two FRBs have been seen to repeat \citep{spitler2016repeating,ChimeRepeat19}. The repeatability of some sources raises the possibility that all FRBs can repeat or suggests that there are at least two classes of FRBs within the currently observed sample \citep[e.g.][]{Palaniswamy18,Ravi18,connor18}.
\frb\ \citep{spitler2014fast} is unique in that it was the first FRB source with repeated bursts detected \citep{spitler2016repeating,scholz2016repeating} and is therefore the most extensively monitored FRB source to date.

The ability to observe multiple bursts from \frb\ permits unprecedented studies of an FRB source and its environment. In particular, \frb\ was localized to 100\,milliarcsecond precision \citep{chatterjee2017direct}; a  low-mass, low-metallicity dwarf host galaxy with a star-forming region was identified at $z\approx0.19$ \citep{tendulkar17,bassa2017frb,kokubo17}; the bursting source was associated with a compact, persistent radio source offset by $\lesssim$ 40\,pc  \citep{marcote2017repeating}; and an exceptionally high, and variable rotation measure (RM) of $\sim 10^{5}$\,rad\,m$^{-2}$ was measured, pointing to an extreme magneto-ionic environment around the burst source \citep{michilli2018}. The recent discovery of a second repeating source (FRB~180814) by the Canadian Hydrogen  Intensity  Mapping  Experiment (CHIME) has not yet resulted in a precise localization \citep{ChimeRepeat19}. As yet, no other published FRB has been precisely localized and definitively associated with a host galaxy, thereby greatly limiting our understanding of their progenitors.

The apparent \frb\ burst activity changes between observing epochs, with periods of enhanced source activity \citep{scholz2016repeating,Oppermann2018}, though it is unclear whether this means that the source itself is intrinsically more active. While an underlying periodicity between the bursts would be strong evidence for a rotating neutron star progenitor, analysis of the burst arrival times has yet to identify a clear periodicity \citep[e.g.][]{scholz2016repeating,zhang18}. Bursts detected at 1.4\,GHz were not seen in simultaneous 150-MHz observations \citep{houben19}. Optical, X-ray and $\gamma$-ray observations that are contemporaneous with the detections of radio bursts have not found any prompt multi-wavelength counterparts to the bursts themselves, nor is there any detectable persistent X-ray and $\gamma$-ray emission \citep{scholz2017xray,hardy17}. 

The detection of a large sample of bursts from \frb, at observing frequencies ranging from 1 to 8\,GHz (e.g.  \citealt{spitler2016repeating,scholz2016repeating,law17,michilli2018,gajjar18,spitler18,zhang18,hessels18}), has led to a variety of observed spectra.
For instance, they cannot be consistently described by a single spectral index, some bursts exhibit a frequency dependent profile evolution, and burst spectra from \citet{law17} are typically limited to 500-MHz wide Gaussian envelopes. 
Additionally, \cite{hessels18} present a sample of bright bursts showing sub-components in their spectra. 

The physical nature of FRB 121102 and the reason for its variable burst spectrum are the subject of many theoretical models \citep{platts18}.
Recent models have been proposed that involve a neutron star in the immediate vicinity of an accreting massive black hole \citep{PenConnor15,tendulkar17,marcote2017repeating,bassa2017frb,michilli2018,zhang18}. Other possibilities include a millisecond magnetar as the central engine of a powerful supernova remnant (e.g. \citealt{MuraseWindbubble,metzger2017millisecond,Beloborodov17,CaoMagnetar,Nicholl17}).
Extrinsic propagation effects have been invoked to describe the  unusual burst structure and morphology \citep{cordesLensing,Main2018,hessels18}. Plasma lenses in the local environment of \frb\ could collectively create caustics that produce an amplification of the burst brightness in certain frequency bands \citep{clegg1998,cordesLensing}. 
Alternatively, the burst spectra could be intrinsic, for instance originating from maser emission \citep[e.g.][]{waxman17,Beloborodov17,metzger19}. Pulsar giant pulses have also been shown to be poorly described by a single spectral index \citep[e.g.][]{meyers17}. Otherwise, a combination of both intrinsic and extrinsic effects could be at play.

In an effort to understand the emission mechanism and environment of \frb, we continue to collect and investigate its bursts. Here, we present 41 bursts from \frb\ detected at 1.4\,GHz using the Arecibo Observatory in two observations from 2016 September on consecutive days. This sample was selected for our analysis due to the large volume of bursts detected in each observation (18 and 23 bursts, respectively) and the short time between observations (the minimum possible wait time before \frb\ transits Arecibo, i.e. one day). The sample includes all bursts that were found, down to an unprecedentedly low detection threshold (via careful visual inspection), and in combination with its size is therefore the largest sample of 1.4-GHz bursts presented to date. The observations are quasi-contemporaneous with the high signal-to-noise ratio (S/N) sample from \citet{hessels18} and concurrent with the Very Large Array (VLA) sample presented in \citet{law17}.

The bursts we detect are predominantly faint (fluence as low as 0.028 Jy ms) and relatively narrow-band: i.e. they do not extend across the full 580-MHz observing bandwidth and many peak in the observing band and fade into the noise towards higher and lower frequencies. These properties originally led to many of the bursts being identified as radio frequency interference (RFI) in our first-pass search of the data. However, the consistent recurrence of such signals compelled further investigation.  We present an analysis of all bursts detected in the two observations that includes examinations of burst spectra, burst energies and wait times. Additionally, we focus on burst detectability to instruct future FRB observations and searches. Observations and data reduction are described in \textsection\ref{methods}. The bursts are presented in \textsection\ref{results} and our findings are discussed in \textsection\ref{discussion}.

\section{Observations and burst search} \label{methods}
The data were recorded using the 305-m William E. Gordon Telescope at the Arecibo Observatory with the L-Wide receiver\footnote{\url{http://www.naic.edu/~astro/RXstatus/Lwide/Lwide.shtml}} (frequency range $1150-1730$\,MHz; system temperature $T_{\rm sys} \approx 25$\,K; gain $G \approx 10.5$\,K/Jy) and the Puerto-Rican Ultimate Pulsar Processing Instrument (PUPPI) data recording backend, which records eight 100-MHz bands (each with 64
channels)\footnote{\url{http://www.naic.edu/puppi-observing/}}. The 8-bit PUPPI data were sampled with a time resolution of 10.24\,$\upmu$s and each channel spans 1.56\,MHz; these were coherently de-dispersed to DM$=557.0$\,pc\,cm$^{-3}$ during the observation, effectively mitigating intra-channel dispersive smearing to $<8.5\,\upmu$s per unit deviation from the fiducial DM value. Full Stokes information was recorded, however given the source's large RM, linear polarization is washed out at 1.4\,GHz \citep{michilli2018} rendering it undetectable in our data. A search for polarization in the brightest burst from our sample was conducted in \citet{hessels18} and did not yield a detection. Hence, we do not present any polarimetric analysis here, though searches are ongoing to identify potential Faraday conversion effects in the data \citep{vedantham18}.

The data were subbanded and down-sampled using \texttt{psrfits\_subband} before searching for bursts. The frequency channel size was increased to 12.5\,MHz and the data were down-sampled in time to a resolution of 81.92 $\upmu$s. 
The search was performed using \textsc{PRESTO}, a standard software package for pulsar searches \citep{presto}\footnote{\url{https://www.cv.nrao.edu/~sransom/presto/}}.
The standard RFI excision tool \texttt{rfifind} was not applied to the spectra to avoid masking large fractions of data. Instead, we opted to excise RFI pulses at later stages of the search process. We generated de-dispersed timeseries (summed across frequency channels) for DMs in the range of $461-661$\,pc\,cm$^{-3}$ in steps of 1\,pc\,cm$^{-3}$, using \texttt{prepsubband}. The resulting timeseries were searched for single pulses using \texttt{single\_pulse\_search.py} to convolve boxcar functions of widths ranging from $81.92$\,$\upmu$s to $24.576$\,ms, or, equivalently, 1 to 300 time bins. Events from each timeseries with $\text{S/N} > 6$ were grouped into astrophysical burst candidates, which were subsequently excluded if the group's peak $\text{S/N} < 8$, and filtered for RFI using the routine presented by \citet{SpS}\footnote{The Single-pulse Searcher code \citep{SpScode} is available at \url{http://ascl.net/1806.013}.}. The recorded times of the remaining burst candidates were used to extract segments of time and frequency data (known as dynamic spectra) from the subbanded data de-dispersed to the DM at which the candidate peaked in S/N. Finally, a diagnostic plot was created for each candidate, containing the de-dispersed dynamic spectrum, burst profile, and relevant meta data.  The diagnostic plots (43 for Observation 1 and 82 for Observation 2) were inspected by eye to evaluate whether the candidates were astrophysical in nature and each was assigned a ranking: RFI, maybe real, definitely real. Judgments were based on what we knew FRBs to look like at the time (late 2016/early 2017), however the ``maybe'' ranking was included to avoid missing interesting and potentially recurring signals in the observations. 
Relevant factors considered in the judgment process include whether the peak DM value was reasonably close to the value reported in earlier studies of \frb\ bursts \citep{spitler2016repeating,scholz2016repeating} and whether the burst extended across most of the frequency band (this was before \citet{hessels18} established that the bursts can appear quite narrow-band at 1.4 GHz).

The data come from the ongoing monitoring of \frb\ (Arecibo program P3054; PI L.~Spitler). We present data from two observations taken on 2016 September 13/09:47:07 and 14/09:50:12, each reported in topocentric UT and lasting 5967\,s and 5545\,s, respectively. One of the bursts (B1) was previously reported in \citet{hessels18}.

The narrow-bandedness, faintness, and, in some cases, relatively large widths (up to 13.5\,ms) of the bursts presented here make the DM determination less precise. A DM value of $560.5$\,pc\,cm$^{-3}$ was established for the high S/N bursts in \citet{hessels18} for which the temporal features are well-resolved. Given that our sample is from the same epoch, we apply the same value to the bursts in this study and find this to work reasonably well.


\begin{figure*}[!hb]
\gridline{\fig{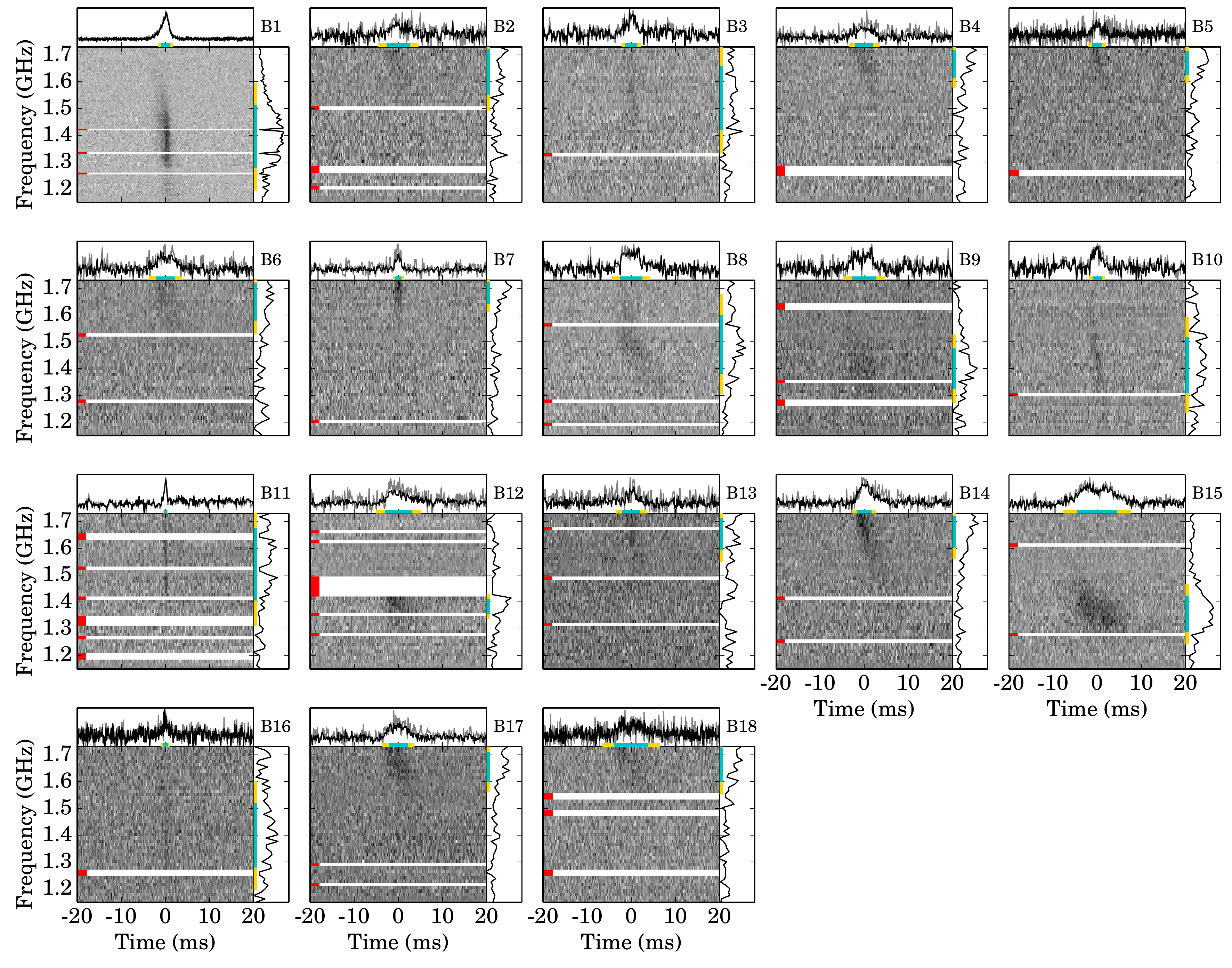}{\textwidth}{(a)}}
\caption{(a) Dynamic spectra of the bursts detected in Observation 1 (2016 September 13), ordered by burst arrival time and de-dispersed using DM$=560.5$\,pc\,cm$^{-3}$. The band-averaged burst profiles (summed in frequency) are shown in the top sub-panels, and the spectra (summed in time across the bursts) are shown to the right of each burst. Each burst signal was fit with a 2D Gaussian in order to determine its bandwidth and duration. The cyan bars extend over the FWHM and the yellow bars extend to the 2$\sigma$ point of each fit. The burst profile/spectrum is obtained by summing frequency/time data within the yellow bars. The burst profile obtained by summing over the full frequency band is shown in grey for comparison, and is typically noisier. Solid white lines are artifacts from frequency channels and time bins that were removed because of RFI contamination, and are marked with red notches.  
\label{fig:DS1}}
\end{figure*}
\begin{figure*}[!p]
\ContinuedFloat
\gridline{\fig{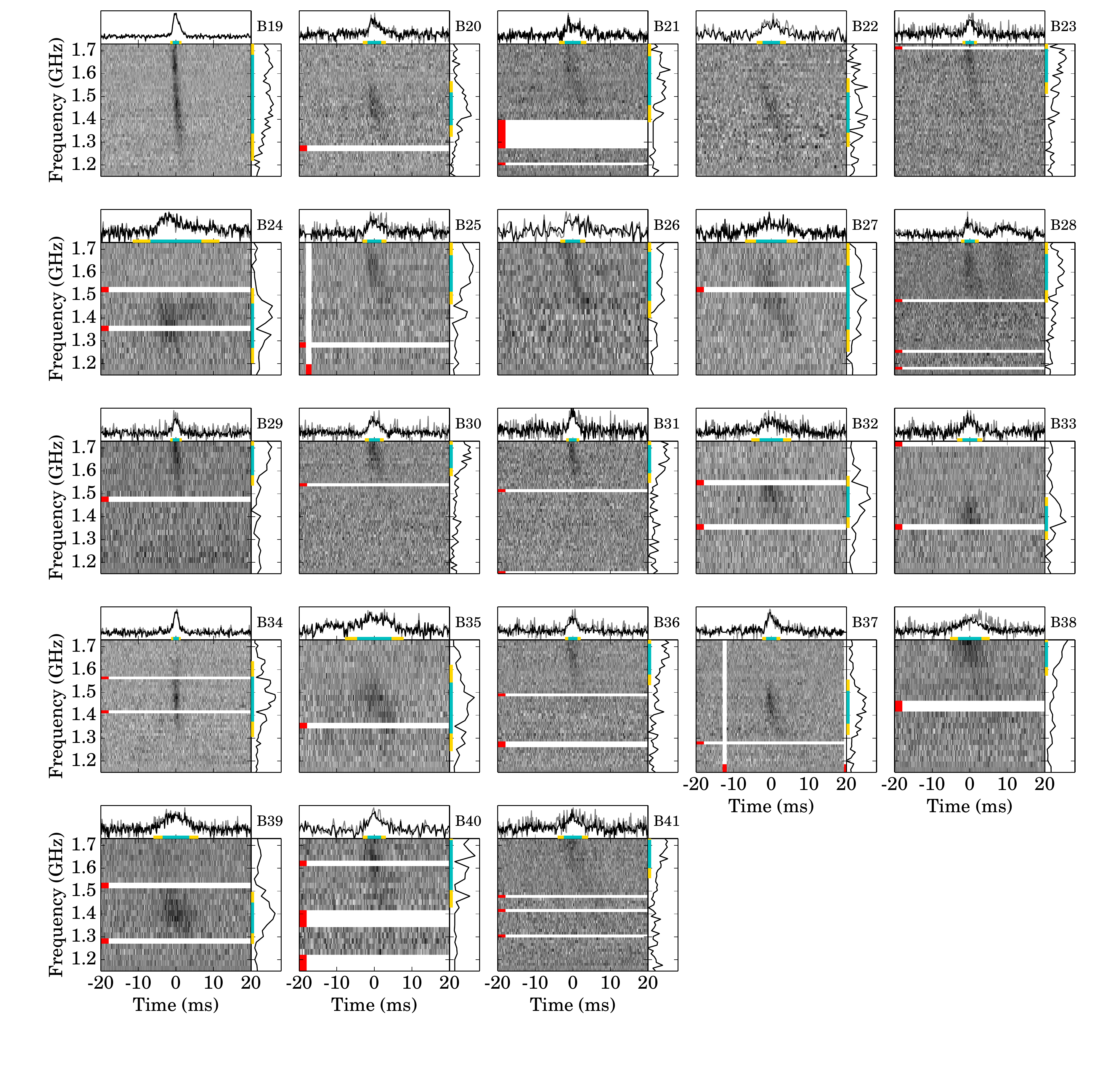}{\textwidth}{(b)}}
\caption{(b) Dynamic burst spectra for Observation 2 (2016 September 14). The bursts have been downsampled in time and downfactored in frequency resolution by differing factors in order to make the burst signals more visible. 
Resolutions of 163.84\,$\upmu$s and 12.48\,MHz: B2-4, B6-15, B17, B19-21, B23, B28, B30-1, B34, B36-7, B40;
81.92\,$\upmu$s and 12.48\,MHz: B5, B16, B18;
163.84\,$\upmu$s and 24.96\,MHz: B24-25, B27, B29, B32-3, B35, B38-9;
327.68\,$\upmu$s and 12.48\,MHz: B22;
327.68\,$\upmu$s and 24.96\,MHz: B26, B40; and 40.96\,$\upmu$s and 6.24\,MHz for B1.
\label{fig:DS2}}
\end{figure*}

\section{Results} \label{results}
\begin{figure*}[!t]
\includegraphics[scale=0.477]{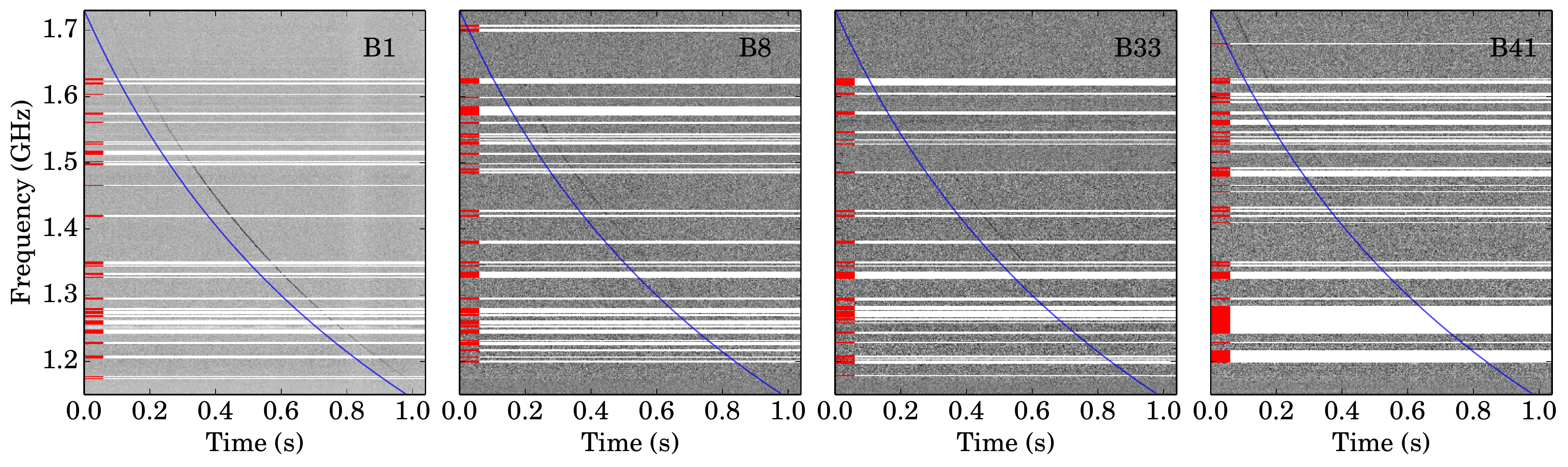}
\caption{Example burst dynamic spectra without correcting for dispersion. The time and frequency resolutions are 2.97\,ms and 1.56\,MHz, respectively. The expected dispersive sweep at DM$=560.5$\,pc\,cm$^{-3}$ is shown in blue for comparison (and is offset in time for clarity). The slight deviation from the expected sweep at later times for B8 is due to the ostensible sub-burst (see \ref{disc:spectra}) visible in the de-dispersed version shown in Figure \ref{fig:DS1}. Solid white lines are artifacts from frequency channels and time bins that were removed due to RFI contamination. These are also marked with red notches. }
\label{fig:dispersed}
\end{figure*}

Dynamic spectra for the 18 bursts detected in the first observation (Observation 1) and for the 23 bursts detected in the second observation (Observation 2) are presented in Figure \ref{fig:DS1}. While bursts B11 and B19 look similar to FRBs presented elsewhere (though see \citealp{shannon18}), many burst signals can be characterized as narrow-band, spanning on average less than a third of the full 580-MHz observing bandwidth (full width at half maximum values, FWHM) with many peaking in brightness within the observing band.  To confirm the bursts' astrophysical nature, we provide some examples without dispersion correction alongside the expected dispersive sweep at 560.5\,pc\,cm$^{-3}$ in Figure \ref{fig:dispersed}. There is an apparent 5.3(5)\,ms wide second burst in the dynamic spectrum of B28, $\sim9$\,ms after the primary burst; it is unclear whether these are two unique bursts or a single, double-peaked burst.

\begin{figure*}[h]
\gridline{\fig{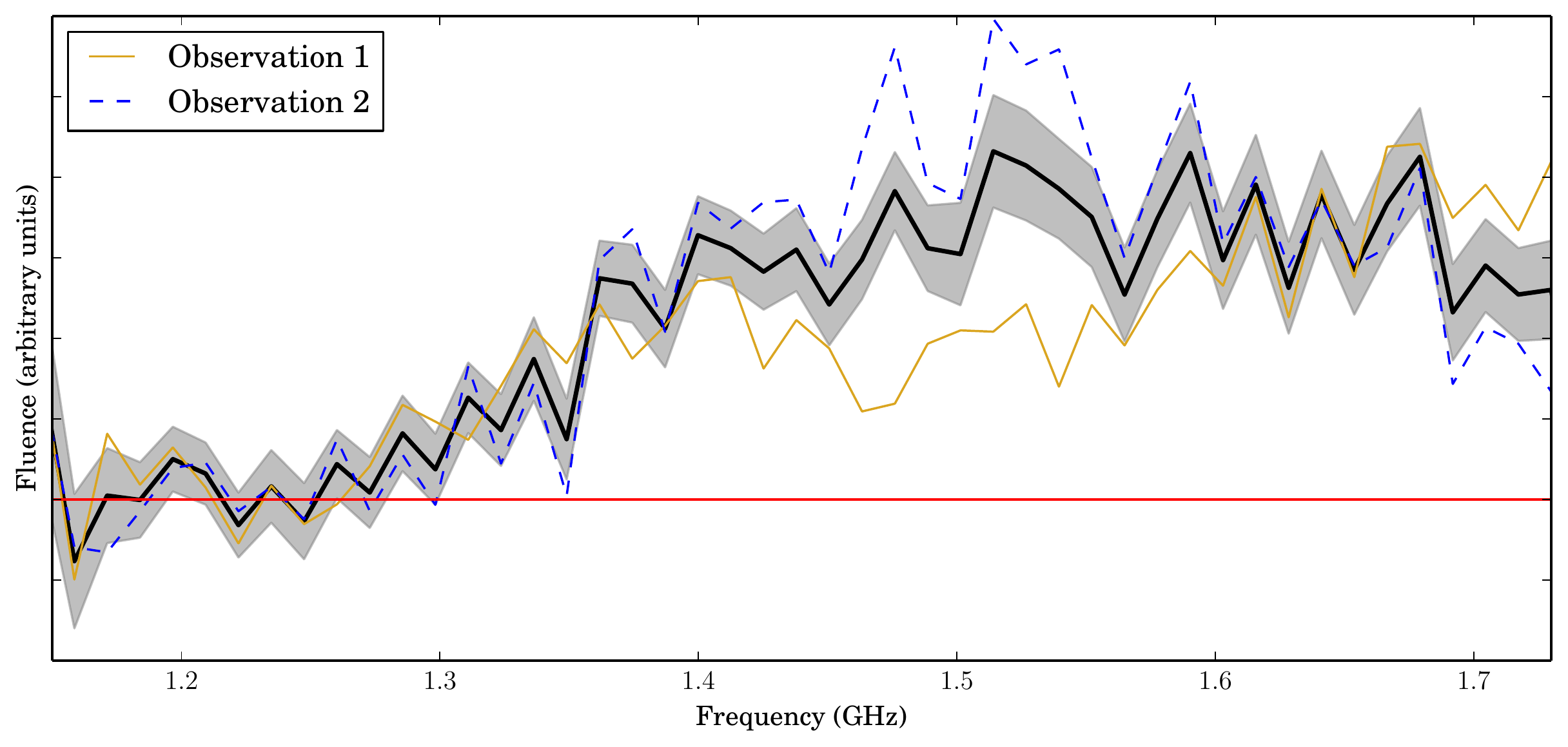}{\textwidth}{}}
\gridline{\fig{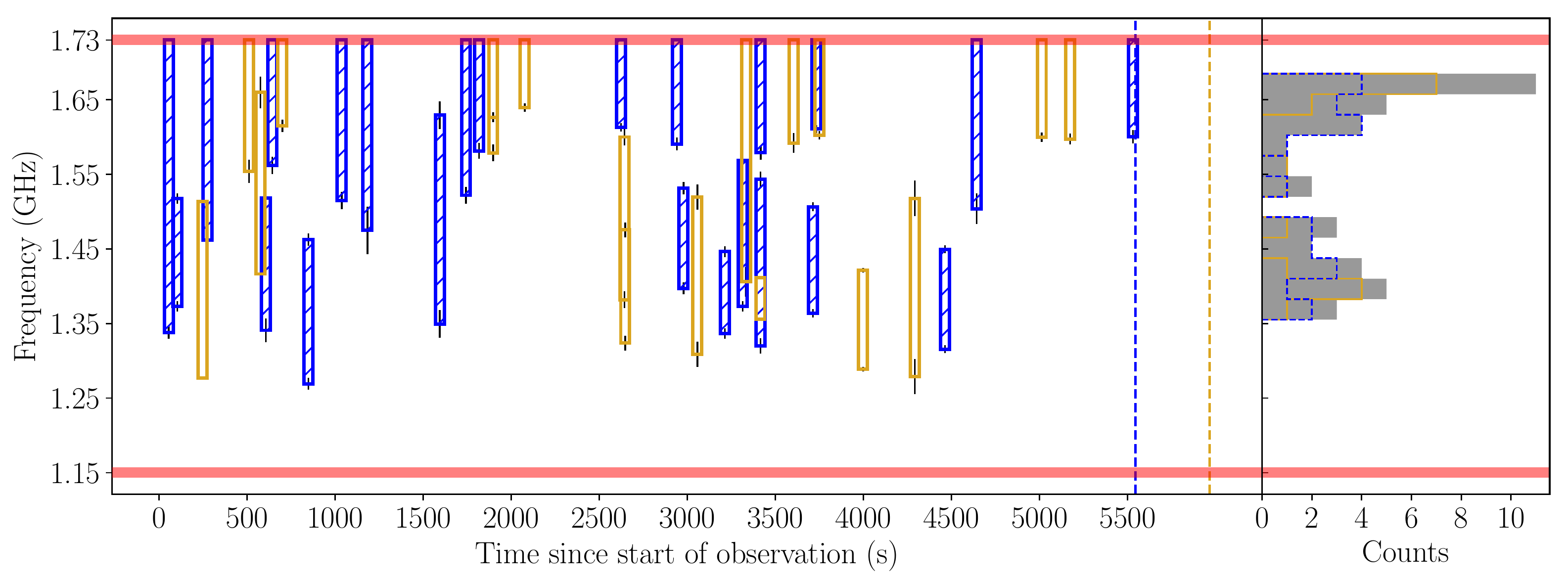}{\textwidth}{}}
\centering
\caption{Top: average burst spectrum, weighted by burst S/N for all bursts (black), Observation 1 (yellow) and Observation 2 (dashed blue). The spectra were bandpass corrected, noise subtracted and normalized before averaging. Errors corresponding to the rms noise fluctuations are shown for the total average spectrum, in grey. The red line marks the normalized baseline.
Bottom, left: burst spectrum (FWHM) as a function of burst arrival time. Dashed vertical lines denote the end time of the observation. Observation 1 is represented by empty yellow markers and Observation 2 by hatched blue  markers. Bottom, right: histogram of burst spectrum centroids (all bursts shown in grey, Observation 1 in yellow and Observation 2 in dashed blue). Red bars indicate the edges of the observing band.
\label{fig:CF}}
\end{figure*}

The properties of each burst are summarized in tabular form in the Appendix (Table \ref{tab:props}). We measure time and frequency peaks and FWHM values using a 2-dimensional Gaussian fit \citep{hessels18}. The intrinsic burst durations range from $0.7-13.5$\,ms and are on average 4.2\,ms. These values are consistent with previous detections at 1.4\,GHz \citep[e.g.][]{spitler2016repeating,scholz2016repeating,hardy17}. We quantify the narrow-band nature of each burst by reporting the burst edges ($f_{high}$ and $f_{low}$, FWHM). Many burst spectra extend beyond the top of the band, and in this case we report the top of the observing band (1730\,MHz). The top panel of Figure \ref{fig:CF} shows the average burst spectrum weighted by the band-averaged burst S/N and corrected for bandpass variations. The burst spectrum as a function of arrival time is shown in the bottom panel. The midpoint of each burst's frequency extent is represented in a histogram in a panel to the right. Collectively, Figure \ref{fig:CF} shows a dearth of bursts below 1.35\,GHz and suggests preferred burst frequencies at this epoch, possibly clustered in time as well, particularly in Observation 1. 

\begin{figure}[h]
\includegraphics[scale=0.405]{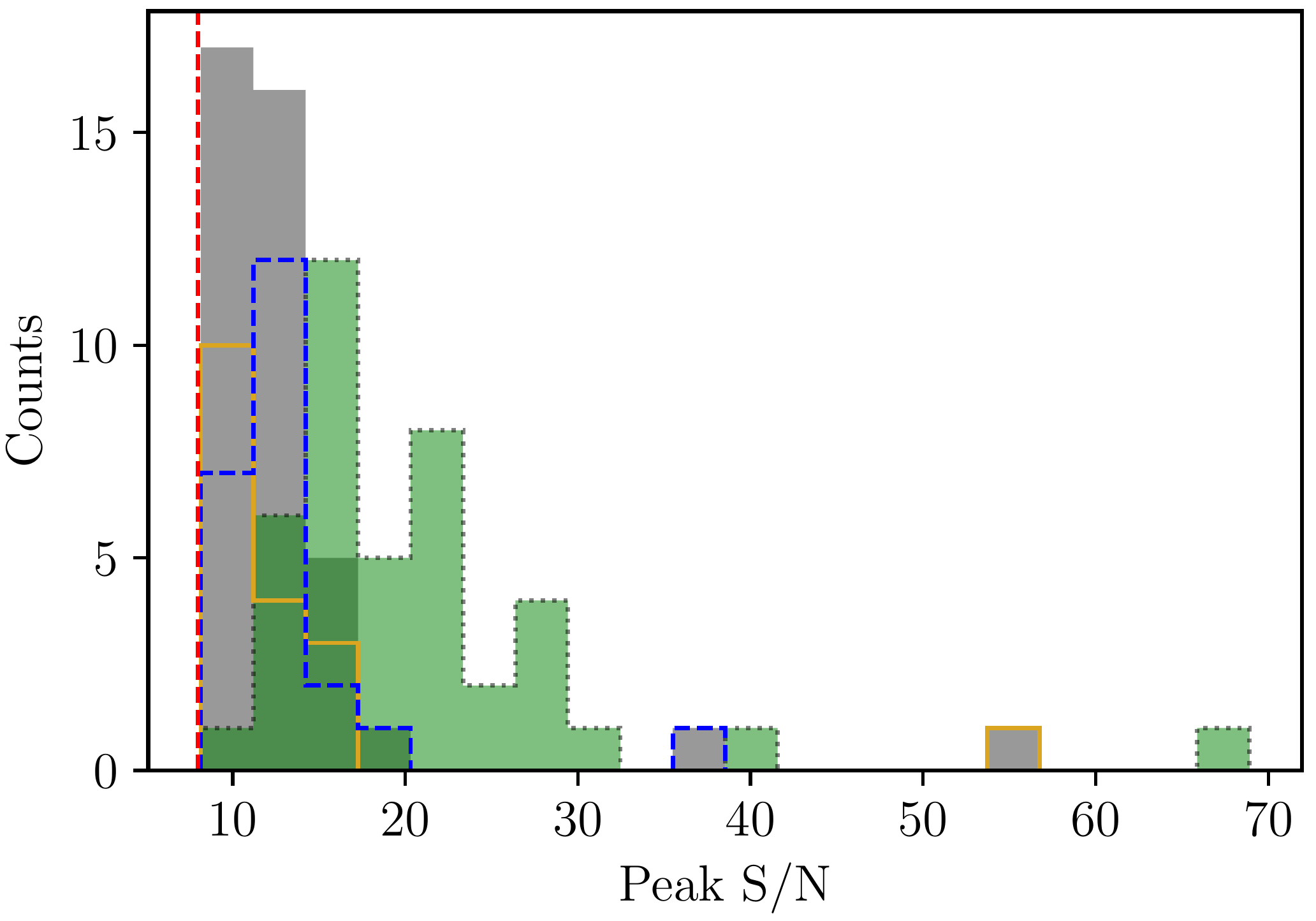}
\caption{Distribution of peak signal-to-noise (S/N) value at which each burst was detected (i.e. at the DM that maximizes S/N), with the search threshold S/N = 8 represented by a dashed red line. Individual distributions are shown in solid yellow (Observation 1) and dashed blue (Observation 2). The combined distribution is shown in grey. The peak S/N, scaled using only frequency channels that contain burst signal (corresponding to the 2$\sigma$ spread of the Gaussian fit shown in Figure \ref{fig:DS1}) according to Equation \ref{eq:snr_mod}, is represented in green, within the dotted contours.
\label{fig:props}}
\end{figure}

The distribution of burst peak S/N (summed over the 580-MHz observing bandwidth and at the DM that maximizes S/N) is reported in Figure \ref{fig:props}. Most bursts were detected just above the detection threshold.
In green we show the scaled-up S/N values obtained using only frequency channels that contain burst signal, as opposed to the full observing bandwidth, since S/N is a function of burst bandwidth.

\begin{figure*}[h]
\gridline{\fig{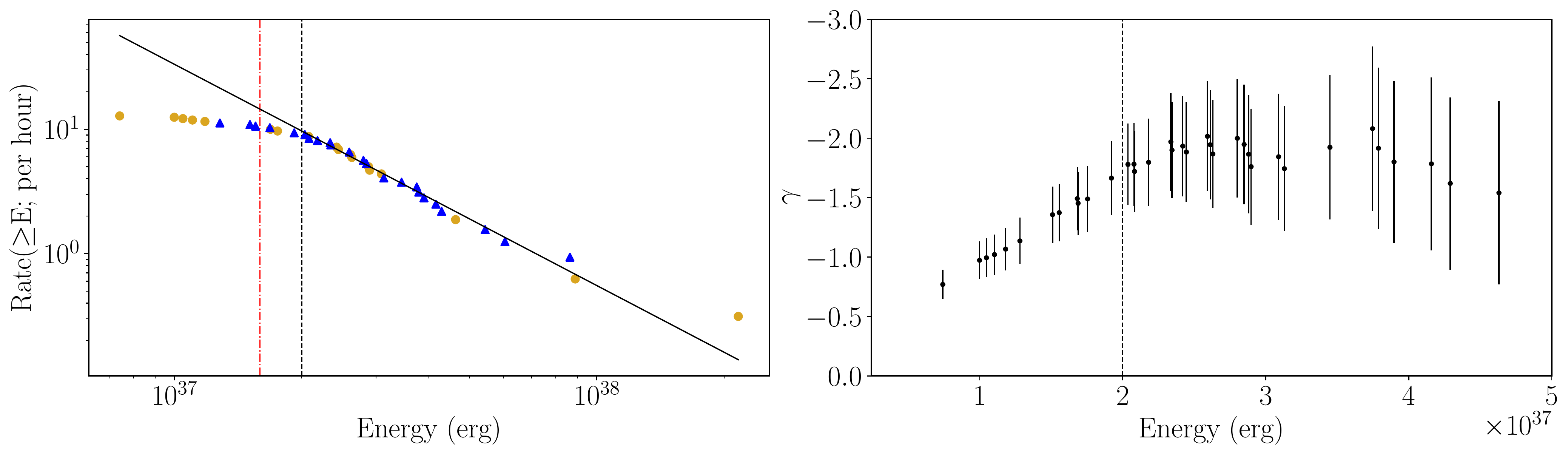}{\textwidth}{}}
\caption{
Left: cumulative distribution of burst energy shown with a power law with $\gamma=-1.8\pm0.3$. Black vertical dashed line denotes the corresponding completeness threshold ($E_{threshold}=2\times10^{37}$) applied. Red vertical dash-dotted line corresponds to the instrument sensitivity threshold for bursts with duration 4.2\,ms and bandwidth 175\,MHz (the average values from our sample). All bursts are considered to have a conservative fractional error of 30\% on the measured fluence value from which energies are derived (Equation \ref{eq:energy}). Right: the power law slope ($\gamma$) determined via Maximum-likelihood estimation as a function of $E_{threshold}$ (each data point corresponds to a successive burst energy).  
\label{fig:energies}}
\end{figure*}

The cumulative distribution of isotropic burst energies is shown in Figure \ref{fig:energies}. We calculate isotropic energy, $E$, matching \citet{law17}:
\begin{align}
E = F\,(\text{Jy\,s}) \times \text{BW (Hz)} \times10^{-23}\mathrm{erg s^{-1}cm^{-2}Hz^{-1}}\notag \\
    \mathrm{\times 4\pi \times L^{2}} \,,
\label{eq:energy}
\end{align}
where $F$ is fluence,  BW is bandwidth (both as reported in Table \ref{tab:props}), and $L$ is the luminosity distance of \frb, 972\,Mpc \citep{tendulkar17}.
We calculate the fluence of a burst over its FWHM duration and FWHM bandwidth. It is calculated by summing across frequency channels that contain the burst to create a timeseries, normalizing a 42-ms time window containing the burst, and converting the signal in each time bin within the FWHM into Jy units using the radiometer equation, as described in Equation 7.12 of \citet{pulsarhandbook}. The normalization step involves defining an off-pulse region and therefore different time resolutions (see caption of Figure \ref{fig:DS1}) are used depending on burst S/N. We consider a conservative fractional error of 30\% for the derived fluence values. Energies are inevitably underestimated by unknown amounts for bursts that extend up to the observing band edge. 
The apparent turnover at lower energies in Figure \ref{fig:energies}, is likely a reflection of bursts being detected close to the sensitivity limit ($\sim1.6\times10^{37}$\,erg for bursts that have the average duration and bandwidth values found for our sample of 4.2\,ms and 175\,MHz, respectively). We use the method of Maximum-likelihood to estimate the slope, $\gamma$, of a power law fit ($R\propto E^{\gamma}$, where R is the rate of bursts with energy $\geq E$ per hour) for various completeness thresholds $E_{threshold}$ \citep[e.g.][]{crawford70,james19}. The black vertical line drawn at $E_{threshold}=2\times10^{37}$\,erg marks the completeness value that we use in the cumulative energy distribution, and was chosen because it is consistent with both the observed turnover and aforementioned sensitivity limit, and is where the distribution of $\gamma$ as a function of $E_{threshold}$ flattens out. Omitting bursts that fall below $E_{threshold}$, we find $\gamma=-1.8\pm0.3$ (and by extension $\frac{dR}{dE}\propto E^{-2.8}$) for the combined sample of bursts and $\gamma=-1.6\pm0.5$ and $\gamma=-1.9\pm0.5$ for Observation 1 and 2, respectively. 

\begin{figure*}[h]
\includegraphics[scale=0.75]{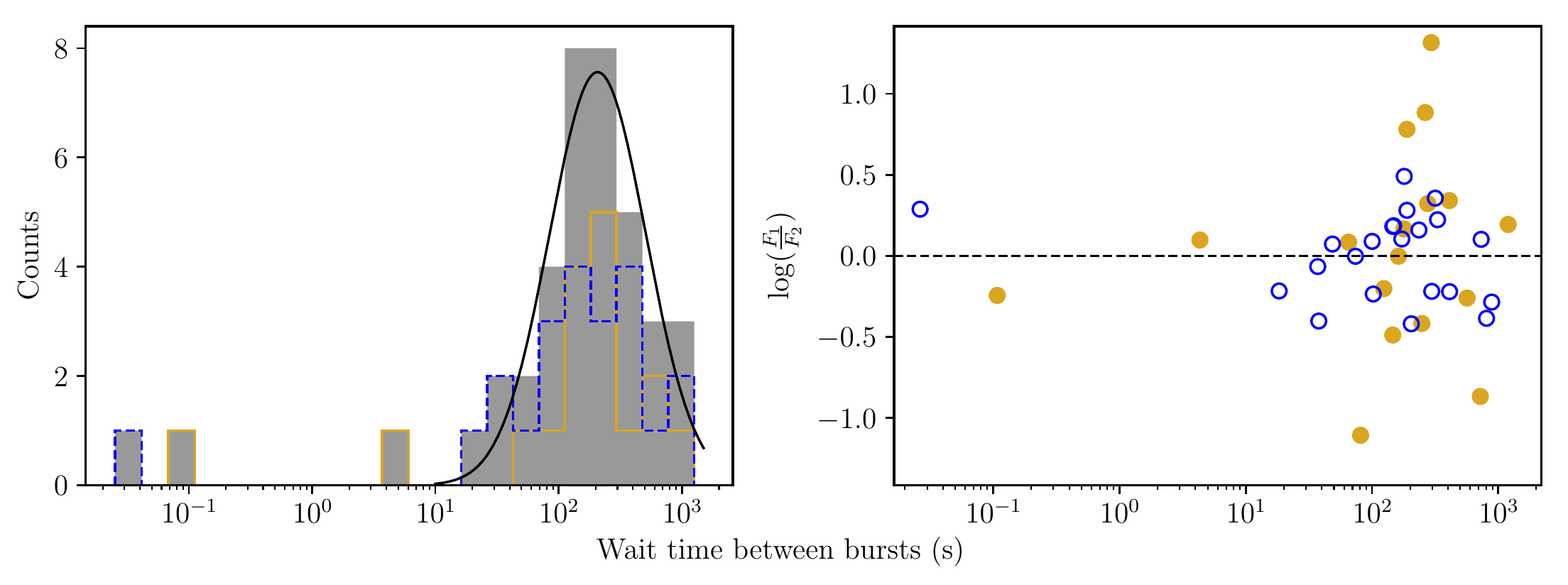}
\centering
\caption{Left: distribution of wait time between consecutive bursts. The distribution of combined wait times for both observations is shown in grey, and the distribution for individual observations is shown in solid yellow (Observation 1) and dashed blue (Observation 2). The combined distribution is fit with a log-normal function centered at $207 \pm1$\,s, omitting the three smallest values. Right: logarithm of the flux ratios in pairs of consecutive bursts as a function of their wait time. The ratio is taken as the fluence of the first burst ($F_{1}$) to that of the second ($F_{2}$). Markers for Observation 1 are yellow and filled, and markers for Observation 2 are blue and hollow. A dashed line is shown along the zero value to show the turnover between brighter first bursts and brighter second bursts. 
\label{fig:wait}}
\end{figure*}

The distribution of burst wait times is shown in the left panel of Figure \ref{fig:wait} and is roughly consistent with a log-normal distribution centered at $207 \pm1$\,s for wait times greater than 10\,s (p-value = 0.73; note the $\sim1.5$\, hour duration of each observation limits our ability to see wait times on the order of $\sim1000$\,s or greater). There is a separate group of wait times below 1\,s. We do not see a relationship between the brightnesses of consecutive bursts and their wait times (right panel of Figure \ref{fig:wait}).  A simple periodicity search was carried out using \textsc{PRESTO}'s \texttt{rrat\_period}, which does a brute force search for a greatest common denominator of the intervals between bursts, given a list of burst arrival times and some trial period. We searched up to spin frequencies of 200\,Hz (corresponding to 5\,ms, which is roughly the average burst width), but did not find consistent common denominator values. This method, however, only works if all bursts have nearly the same rotational phase. More sophisticated periodicity searches are ongoing.

\section{Discussion}\label{discussion}
High resolution data, known source DM, and a rigorous burst search process, have allowed us to approach the theoretical detection threshold of the telescope (despite strong RFI) and probe low burst energies. For instance, \citet{scholz2016repeating} used a S/N cut of $\sim12$ and would have missed many of the bursts presented here. 
In \textsection\ref{disc:spectra}, we discuss the burst spectra. In \textsection\ref{disc:distributions}, we discuss the distributions of burst energies and wait times. Finally, we show how our findings can inform the research community's search strategy for new FRB sources and repeat bursts in \textsection\ref{disc:detection}.
 
\subsection{Burst spectra}
\label{disc:spectra}
In \textsection\ref{methods} we outlined our burst search method, which included manual candidate classification. Notably, many of the bursts presented in this study were initially placed in either the ``maybe real'' category or RFI; FRBs were not known to be narrow-band at the time of the search. The burst candidates were later promoted to ``definitely real'' after more careful consideration of the candidates as a whole and in particular their recurrence.

\begin{figure*}[t]
\includegraphics[scale=0.4]{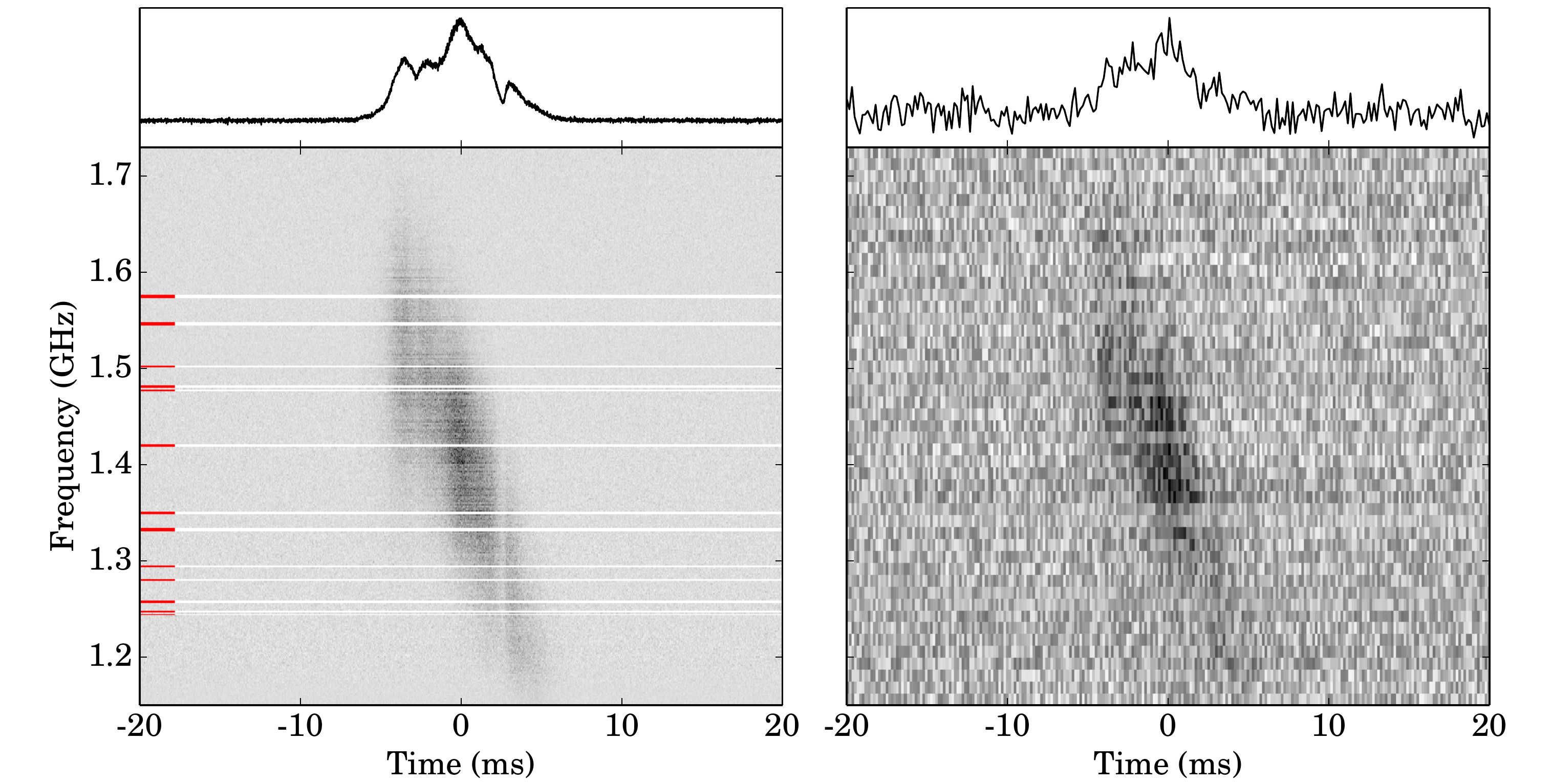}
\centering
\caption{Left: \frb\ multi-component burst AO5 from \cite{hessels18}, de-dispersed at 560.5\,pc\,cm$^{-3}$ and a time and frequency resolution of 10.24\,$\upmu$s and 1.56\,MHz respectively. Right: the same burst after adding noise until the burst S/N is comparable to the values of the bursts we present here. The spectrum was then downfactored and downsampled to be comparable to the bursts shown in Figure \ref{fig:DS1}. \label{fig:noisy}}
\end{figure*}

In an attempt to understand this subset of bursts in the context of other \frb\ bursts, we compare to the multi-component bursts presented in \citet{hessels18} (see Figure \ref{fig:noisy}). The bursts presented in \citet{hessels18} were chosen strictly for their high S/N. The sub-bursts contained in the full multi-component burst envelope 
emit with a characteristic bandwidth of $\sim$250\,MHz at $1.4$\,GHz and envelopes are as large as $\sim$10 ms wide \citep{hessels18}. The sub-bursts drift down in frequency during the duration of the burst and the leading edges of sub-bursts are often sharper than the trailing edges. Similarly, some bursts from our sample tend to lower frequencies with time, and some also fade towards the trailing edge (e.g. B12, B14, B18, B20, B37, B41). Our bursts typically have durations within the observed multi-component burst envelope sizes of the \citet{hessels18} sample. \citet{hessels18} argue that the narrow-bandedness cannot be from propagation effects in the Milky Way. Similar spectral behaviour has been observed in bursts from FRB~180814 \citep{ChimeRepeat19}.

We explore the possibility that the narrow-band signals from this study are faint multi-component bursts where only the brightest sub-burst(s) and surrounding diffuse emission are detected. To do this, we characterize the noise around the burst to generate a Gaussian noise distribution from which noise is drawn and added to the multi-component burst until a S/N comparable to those of our bursts is reached. The result of this procedure for burst AO5 from \cite{hessels18} is provided in Figure \ref{fig:noisy}. We find the results to convincingly support the hypothesis that, with a more sensitive receiver, our narrow-band bursts might have looked similar to the multi-component bursts, particularly given that the burst profiles (Figure \ref{fig:DS1}) are generally not well described by a single Gaussian profile.

In the last year, it has become increasingly clear that FRBs are not always broadband 
\citep{law17,michilli2018,zhang18,shannon18,hessels18,ChimeRepeat19,Chime19}. Many FRBs have been discovered using relatively narrow-band receivers (e.g. ranging from 16 and 32\,MHz for Molonglo/UTMOST to $\sim336$\,MHz for ASKAP), which makes it difficult to gather much information about the burst's broadband spectrum.

Interestingly, in the 3-GHz VLA observations that are contemporaneous with the Arecibo observations we present here, no bursts were detected during Observation 1 and only one separate burst was detected by VLA during Arecibo Observation 2 \citep{law17}. This VLA burst is not visible in our contemporaneous Arecibo data. However, one burst was simultaneously observed by both instruments a few days after Observation 2, detected across the full L-wide band ($1.15-1.73$\,GHz) and from about 2.5 to 3.2\,GHz at the VLA \citep{law17}.
This collectively tells us about the limited detectable spectral extent of individual bursts and that bursts are detectable at preferred frequencies that change from epoch to epoch on $\sim$day timescales.

Taking into account the lower sensitivity of the VLA observations (0.148\,Jy\,ms at the detection threshold S/N$=7.4$ compared to Arecibo's $\sim0.02$\,Jy\,ms at the S/N$=8$ threshold, both corresponding to bursts detectable across the full band) and broader observing bandwidth (1024\,MHz), a burst reaching the VLA at a frequency $\sim3$\,GHz and having the average characteristics of the reported sample would be detectable with the VLA if it has approximately $F>0.49$\,Jy\,ms (see Equation \ref{eq:snr_mod} in \textsection{\ref{disc:detection}}) or, equivalently, $E=9.7\times10^{37}$\,erg. Here, we have assumed a temporal burst width equal to the VLA integration time of 5\,ms and a burst spectral width of 307\,MHz, corresponding to our average observing bandwidth fill fraction of 30\%. According to Figure~\ref{fig:energies}, there should be 0.6 bursts with $E>9.7\times10^{37}$ erg per hour and the VLA observations were two hours each.

These wide-band simultaneous observations likely point to a highly variable burst spectrum that can be limited to narrow frequency ranges (tens to hundreds of MHz) per burst. Bursts as narrow in frequency as those presented here have not been clearly detected in higher frequency observations \citep[e.g.][]{hessels18,michilli2018,gajjar18}, though this is consistent with the observed behaviour of apparent increase in sub-burst bandwidth at higher frequencies noted by \cite{hessels18}. The physical reason for this behaviour is unknown.

In a large sample of \frb\ bursts detected within 4 hours, \citet{gajjar18} and \citet{zhang18} found bursts to occur at preferred frequencies in their $4-8$\,GHz band. We show that this behaviour is present at 1.4\,GHz as well (Figure \ref{fig:CF}).
The narrow-band signals with preferential emission frequencies and the dearth of bursts below 1.35\,GHz are consistent with the effects of plasma lensing, which can produce spectral islands due to caustic peaks \citep{cordesLensing}. For a single Gaussian lens, there may be double-peaked gains as a function of frequency, at some epochs. The separations of these gain cusps depend on the offset of the lens from a direct line of sight to the source and is thus epoch dependent, even for the same lens. A reasonable combination of parameters could produce the observed preference in emission frequency in our two observations.  A prediction would be that at another epoch where multiple bursts are seen, the separation of a double cusp would be different. A slightly more complex lens (such as a distorted Gaussian) can show multiple spectral islands and, if there is a population of lensing structures, like filaments in the Crab Nebula \citep[e.g.][]{temim2006}, there can be even more islands. A related prediction is that for observations at different widely spaced frequencies, we would expect spectral islands to be different if seen at all in a separate frequency band. Further simultaneous observations at multiple observing bands are needed to improve our understanding of the spectral behaviour of \frb\/.

We do not rule out the possibility that these bursts are intrinsically narrow-band, for example resulting from maser emission \citep[e.g][]{waxman17,metzger19}. Another possible way to obtain narrow-band signals is if the bursts have an intrinsic frequency dependent brightness, but are inherently weak such that only the brightest portions are detected. It is likely that a combination of both intrinsic and extrinsic mechanisms are producing the observed complex spectra of \frb\ bursts. For instance, lensing may boost burst envelopes at preferred frequencies and the structures within the envelope may be intrinsic. However, there currently are not many emission models that can explain the observed burst structure (though see \citealp{metzger19}).

\subsection{Burst energies and wait times}
\label{disc:distributions}
The cumulative distribution of observed burst energies from our study (Figure \ref{fig:energies}) is consistent with a single powerlaw fit with $\gamma=-1.8\pm0.3$, above $E_{threshold}=2\times10^{37}$\,erg. This value is at odds with the results of \citet{law17}, where $\gamma=-0.7$ was found to consistently describe the cumulative energy distributions of a separate sample of \frb\ bursts detected by the VLA at 3\,GHz, the Green Bank Telescope (GBT) at 2\,GHz and Arecibo (though with the slightly less sensitive ALFA receiver) at 1.4\,GHz. The fluence values of the bursts detected by the latter two telescopes were scaled to energy with the assumption that the full bursts were detected (ALFA has an observing bandwidth of 323\,MHz), which likely underestimates the energy in most cases. A completeness threshold was not applied to their sample.
\citet{law17} suggest that a consistent power law index for observations with different frequencies and detection rates is connected to the underlying emission mechanism. The \citet{law17} analysis includes bursts with energies $>2\times10^{37}$\,Jy\,ms. Therefore, our sample of bursts probes the burst energy distribution of \frb\ to unprecedentedly low energies. 

There are many potential complications to consider in analyzing the distribution of burst energies from \frb\/. First, there are clearly parts of some bursts being missed due to the limited observing bandwidth. Second, there may be fainter burst sub-components that fall below the detection threshold. Third, the presence of extrinsic propagation effects would skew the results of any energy distribution. Therefore, expecting a power law to describe the cumulative energy distribution is likely an over-simplification. Additionally, determining where the sample is complete can have a large effect on the steepness of the slope (Figure \ref{fig:energies}, right).
In any case, possible reasons for the difference in slope of the power law approximation used here and in \citet{law17} can stem in part from the different energy range being sampled for a burst energy distribution that is more complex. 
For instance, 
\cite{Karuppusamy2010} have shown that the Crab pulsar's pulse intensity distribution is multi-modal, peaking at lower intensities (the regular pulsar-like pulses), followed by a log-normal distribution and finally an extended power law tail attributed to the Crab's giant pulses. In contrast, regular pulsars are known to have consistent energy distributions from epoch to epoch, when sampling averaged pulses, and are typically described by a normal or exponential function \citep{hesse74,ritchings76}. Single-pulse emission from both radio magnetars and pulsars, however, tend to have log-normal flux density distributions \citep{levin12,burkespoaler12}.

From the cumulative energy distribution and derived $\gamma$, we can test the hypothesis that the persistent radio source associated with \frb\ is due to the emission of many faint bursts that fall below the detection threshold of our telescopes. We do this by making the simple assumption that the energy distribution is described by $R\propto E^{-1.8}$ (see \textsection\ref{results}) at some minimum energy $E_{min}$, and by setting the known luminosity of the persistent source $L_{p}=3\times10^{38}$\,erg\,s$^{-1}$ \citep{marcote2017repeating} equal to $E_{min}$ times the rate at $E_{min}$. Using our results from Figure \ref{fig:energies} for $E_{threshold}=2\times10^{37}$ erg, we can solve for $E_{min}$ using the following approximation:
\begin{equation}
L_{p} \approx R_{threshold}\Bigg({\frac{E_{min}}{E_{threshold}}}\Bigg)^{\gamma}E_{min}\,,
\label{eq:lum}
\end{equation}
where $R_{threshold}=360^{-1}$\,s$^{-1}$ (the rate at $E_{threshold}$) and $\gamma=-1.8$. The resulting minimum energy $E_{min}=4.3\times10^{32}$\,erg corresponds to a rate of $\sim700$ bursts per millisecond. Therefore, for the assumed power law energy distribution and given that the burst widths in our sample are on the order of a millisecond, it is implausible that the persistent radio emission is generated by a high rate of low energy bursts.

The giant pulse model has been proposed as an emission mechanism for FRBs, where bursts are extreme versions of giant-pulses like those observed in the Crab pulsar \citep{connor2016GP,cordes2016GP,lyutikov2016GP,katzOverview}. Pertinently, the lack of observed correlation between burst wait times and energy (Figure \ref{fig:wait}) is consistent with pulsar giant pulse emission \citep{Karuppusamy2010}. Of relevance to magnetar related models, the log-normal shape of the burst wait time distribution from Figure \ref{fig:wait} is also seen for soft gamma repeaters, which are a type of magnetar \citep{gogus1999,gogus2000,wang17}.

A distinct smaller population of \frb\ burst wait times below 1\,s has also been noted by \citet{katz18a}, \citet{zhang18} and \citet{Li19}. With those wait times omitted, a log-normal function can also reasonably describe the wait time distributions found in both studies, which peak at $\sim75$\,s and $170$\,s respectively. The \citet{Li19} analysis included the $4-8$-GHz \cite{zhang18} bursts, which dominate the sample. Due to a larger sample of bursts, the gap between populations in the wait time distribution of the \citet{zhang18} bursts is slightly smaller (beginning at 600\,ms) relative to ours. With a larger sample of bursts, we might see more wait times fill the observed gap.  The constancy of the distribution of burst arrival time intervals for both of our observing days suggests that the burst detection rate can be consistent on $\sim$day timescales. \citet{Li19} agree with our finding that burst fluence is independent of wait time. Studying burst wait times for (simultaneous) observations at different frequencies could provide additional constraints to the emission mechanism and/or extrinsic propagation effects involved. 

We have uniform sensitivity to wait times between $\sim1000$\,s (on the order of the observation length) down to tens of milliseconds, at which point ambiguities in distinguishing multi-peak bursts from single bursts with small separations in arrival time (e.g. B28) complicate the analysis, as well as periodicity searches. For high S/N bursts, the separation between sub-bursts was found to be $\sim1$\,ms \citep{hessels18}. \citet{zhang18} have reported a burst pair separated by 2.56\,ms which, if both are unique, would be the most closely spaced bursts detected to date. Other ambiguous pairs reported in their analysis are separated by the same order of time as the sub-components of burst B28 ($\sim$10\,ms). Excluding inconclusive cases and assuming B36 from our analysis is a singular burst, it has one of the shortest wait times observed to date at 26\,ms.                                                                              


\subsection{Implications for FRB searches}\label{disc:detection}

The definition of a `canonical' FRB is changing, and this is important for considering which detected signals are of genuine astrophysical origin \citep{foster18}.  We emphasize that the standard pulsar single-pulse search techniques widely used in FRB searches are likely to have missed most of the bursts we present. If not for the development of tailored search algorithms, conservative search filters \citep{SpS}, and human inspection (possible in this case because we are targeting a known source with known DM), the tally of Arecibo bursts from \frb\ would be reduced by about one third. Important to keep in mind is that our observations benefit from high frequency and time resolutions and the source DM was previously known. Assuming there has not been a significant change in the activity level of these bursts, signals like the ones we present here have likely been missed in previous \frb\ observations presented in for example \citet{spitler2016repeating} and \citet{scholz2016repeating}. 

The features that set many of the bursts presented in this study apart from the other bursts observed from \frb\ and other FRB sources are their combined narrow bandwidth and faintness. In the rest of this subsection, we discuss the associated detection implications and suggest possible solutions.

The search techniques used to generate our burst candidates involve consideration of the peak S/N, obtained after summing all de-dispersed frequency channels. A burst's S/N depends on multiple factors \citep{cordes2003searches}, including the width and the intrinsic fluence of the burst, which is defined as the area of the burst (i.e. the amplitude after adding signal across the frequency band, multiplied by burst width). S/N scales with fluence, $F$, and width, $w$, as
\begin{equation}
S/N \propto \frac{F}{\sqrt{w}}\, .
\label{eq:snr}
\end{equation}
Thus, narrower bursts are more easily discerned from the noise than wider bursts of equivalent fluence.
A burst's limited bandwidth will also contribute significantly in lowering the peak S/N, as it will be diluted by noisy frequency channels after summing together in frequency to create the timeseries. This effect is demonstrated in Figure \ref{fig:props} and is directly visible by comparing the black and grey burst profiles in Figure \ref{fig:DS1}. Therefore, Equation \ref{eq:snr} should be modified to take into account the fraction of the band where signal is detected over the noise level, $\frac{\nu_{signal}}{\nu_{band}}$:
\begin{equation}
S/N \propto \frac{F}{\sqrt{w}}\Bigg(\frac{\nu_{signal}}{\nu_{band}}\Bigg)\, .
\label{eq:snr_mod}
\end{equation}
It is very likely that FRB signals that do not fill the entire observing bandwidth are being missed. In this study, we have shown that the three aforementioned properties (faint, narrow in frequency and wide in time) that can reduce detectability often overlap, compounding the difficulty of detecting such signals.

Deviations from the true source DM in the de-dispersed timeseries reduces the peak S/N value in a directly proportional fashion \citep{cordes2003searches}. Bursts that are narrow in frequency, large in temporal width and faint will contribute to uncertainties in the DM measurement, regardless of the method of determination used (e.g. visual dynamic spectrum alignment or peak S/N maximization). Furthermore, de-dispersed timeseries at a wide range of DMs constitute a fundamental aspect of single pulse searches, especially if the source DM is unknown. As described in \textsection{\ref{methods}}, events found in each timeseries are grouped into astrophysical candidates. It is in this crucial grouping step where bursts similar to those presented in this analysis can be missed. According to \cite{cordes2003searches}, bursts narrower in time will peak more sharply in their distribution of S/N as a function of DM, causing them to be easier to find. Therefore, search algorithms sensitive to slow peak S/N turnovers in the timeseries are necessary to detect wider bursts that are also faint. The challenge is compounded for narrow-band bursts, as their S/N versus DM distribution will be similar to that of narrow-band RFI.

Burst candidate classifiers usually consider broad-band bursts. It could be important to change this aspect, despite the associated difficulty in distinguishing between real bursts and narrow-band RFI, as the bursts become progressively narrow-band. For repeating FRBs, more weight could be given to a burst candidate if others have been seen to peak at that frequency. Especially with the advent of new telescopes with larger fractional observing bandwidths (e.g. \citealp{chimeCollab}), observers may wish to consider the effects of peak S/N dilution in the case of narrow-band bursts. A possible solution is to apply matched filtering techniques in the frequency domain, though this will be computationally costly and increase the number of candidates \citep{zhang18}.


\section{Conclusions}
We have presented and analyzed 41 \frb\ bursts resulting from two consecutive 1.4-GHz observations at the Arecibo Observatory and rigorous burst search methods. Our analysis has probed the faintest bursts from \frb\ in a period of high burst detection rate. We have shown the bursts to be detectable at preferred frequency ranges that vary between epochs on timescales of $\sim$days, which is expected if plasma lensing is at play. Additionally, we have demonstrated that we have likely observed faint versions of previously reported bursts showing complex structure.
We have found a power law fit to the cumulative burst energy distribution to be at odds with previously reported slope values and have discussed possible reasons for the discrepancy. We have placed constraints on the idea that the persistent radio source associated with \frb\ is from a high rate of low energy bursts. We have found the wait time between bursts to follow a log-normal distribution, which has also been observed in previous \frb\/ studies as well as some magnetars. We have identified a sub-group of bursts with wait times below 1\,s which is consistent with previous reports. 
The faint and narrow-band bursts we have presented bolster the findings of recent studies that show that FRBs are not always detectable across the full observing band. We have discussed the challenges associated with detecting such signals and have provided recommendations for future FRB searches to minimize the likelihood of missing new FRBs and possible repeat bursts.


\acknowledgments
We thank A.M. Archibald, B. Margalit, K. Nimmo, J.M. Weisberg and the referee for their helpful comments and insight. K.G., D.M. and J.W.T.H. acknowledge funding from an NWO Vidi fellowship and from the European Research Council under the European Union's Seventh Framework Programme (FP/2007-2013) / ERC Starting Grant agreement nr. 337062 (``DRAGNET''; PI: Hessels). L.G.S. acknowledges funding from the Max Planck Society. JMC and SC acknowledge support from the National Science Foundation (AAG 1815242).

\vspace{5mm}
\facilities{Arecibo} 
\software{Astropy \citep{astropy1,astropy2}, PRESTO \citep{presto}, PSRCHIVE \citep{psrchive}, psrfits\_utils\footnote{\url{https://github.com/demorest/psrfits\_utils}}}

\newpage
\appendix
\section{Burst properties}
\begin{longtable}{llllllll}
\caption{Burst properties.
\newline
\footnotesize{$^{a}$Arrival time of burst peak at the solar system barycenter, after correcting to infinite frequency using a DM$=560.5$\, pc\,cm$^{-3}$ (value determined by \citealp{hessels18}).}\\
\footnotesize{$^{b}$FWHM. The top edge of the band (1730\,MHz) is reported for bursts that run over the top of the band.}\\
\footnotesize{$^{c}$Given for the DM that maximizes peak S/N}\\
\footnotesize{$^{d}$A conservative 30\% fractional error is assumed.}\\
\footnotesize{$^{e}$Values correspond exclusively to main burst (component).}
\label{tab:props}}\\

\toprule
Burst ID & Peak Time (MJD)$^{a}$ & Wait time (s) & Width (ms)$^{b}$ & $f_{high}$ (MHz)$^{b}$ & $f_{low}$ (MHz)$^{b}$ & S/N$^{c}$ & Fluence (Jy\,ms)$^{d}$ \\
\midrule
\endfirsthead
\multicolumn{8}{l}{\tablename\ \thetable\ -- \textit{Continued from previous page}} \\
\hline
Burst ID & Peak Time (MJD)$^{a}$ & Wait time (s) & Width (ms)$^{b}$ & $f_{high}$ (MHz)$^{b}$ & $f_{low}$ (MHz)$^{b}$ & S/N$^{c}$ & Fluence (Jy\,ms)$^{d}$ \\
\hline
\endhead
\hline \multicolumn{8}{l}{\textit{Continued on next page}} \\
\endfoot
\hline
\endlastfoot
B1 & 57644.411070948459 & --- & 1.99(1) & 1514(2) & 1277(2) & 57 & 0.8\\
B2 & 57644.414122641494 & 263.666 & 5.4(5) & 1730 & 1554(16) & 8 & 0.11\\
B3 & 57644.414877772811 & 65.243 & 2.6(2) & 1660(21) & 1416(21) & 9 & 0.09\\
B4 & 57644.416313742695 & 124.068 & 4.2(4) & 1730 & 1615(8) & 9 & 0.14\\
B5 & 57644.430169165447 & 1197.107 & 2.4(2) & 1730 & 1626(7) & 10 & 0.09\\
B6 & 57644.430170411426 & 0.108 & 4.4(4) & 1730 & 1579(11) & 10 & 0.16\\
B7 & 57644.432241693663 & 178.959 & 1.5(1) & 1730 & 1640(6) & 13 & 0.11\\
B8 & 57644.438793986075 & 566.118 & 5.1(3) & 1600(11) & 1382(11) & 14 & 0.19\\
B9 & 57644.438844194447 & 4.338 & 5.6(4) & 1476(10) & 1324(10) & 9 & 0.15\\
B10 & 57644.443589025832 & 409.953 & 2.1(2) & 1520(17) & 1309(17) & 11 & 0.07\\
B11 & 57644.446787095105 & 276.313 & 0.73(6) & 1730 & 1406(20) & 12 & 0.03\\
B12 & 57644.447726499660 & 81.164 & 6.0(3) & 1411(2) & 1356(2) & 11 & 0.4\\
B13 & 57644.449914542223 & 189.047 & 2.0(2) & 1730 & 1592(13) & 8 & 0.07\\
B14 & 57644.451604445720 & 146.008 & 3.3(2) & 1730 & 1602(5) & 17 & 0.22\\
B15 & 57644.454476480409 & 248.144 & 9.1(2) & 1421(3) & 1289(3) & 17 & 0.6\\
B16 & 57644.457882214032 & 294.255 & 1.1(1) & 1518(24) & 1279(24) & 10 & 0.028\\
B17 & 57644.466221285402 & 720.495 & 4.2(2) & 1730 & 1599(6) & 14 & 0.20\\
B18 & 57644.468094375407 & 161.835 & 7.7(5) & 1730 & 1597(7) & 12 & 0.21\\
B19 & 57645.411087942470 & --- & 1.78(4) & 1730 & 1338(8) & 37 & 0.20\\
B20 & 57645.411650660666 & 48.619 & 3.7(2) & 1517(7) & 1373(7) & 12 & 0.17\\
B21 & 57645.413643746637 & 172.202 & 4.3(3) & 1730 & 1462(16) & 11 & 0.13\\
B22 & 57645.417466348888 & 330.272 & 4.7(5) & 1518(16) & 1341(16) & 10 & 0.08\\
B23 & 57645.417896457519 & 37.161 & 2.4(2) & 1730 & 1562(12) & 12 & 0.09\\
B24 & 57645.420264948967 & 204.637 & 13.5(6) & 1463(8) & 1269(8) & 11 & 0.24\\
B25 & 57645.422453982079 & 189.132 & 3.8(3) & 1730 & 1515(11) & 12 & 0.13\\
B26 & 57645.424144819131 & 146.09 & 4.0(6) & 1730 & 1475(32) & 10 & 0.08\\
B27 & 57645.428903822678 & 411.176 & 8.2(6) & 1629(19) & 1349(19) & 14 & 0.14\\
B28$^{e}$ & 57645.430621479631 & 148.405 & 2.8(2) & 1730 & 1522(11) & 10 & 0.09\\
B29 & 57645.431477351958 & 73.947 & 1.9(2) & 1730 & 1581(11) & 12 & 0.09\\
B30 & 57645.440813632951 & 806.654 & 3.0(2) & 1730 & 1613(6) & 13 & 0.22\\
B31 & 57645.444479942598 & 316.769 & 2.1(2) & 1730 & 1591(9) & 12 & 0.1\\
B32 & 57645.444918501853 & 37.891 & 6.1(4) & 1531(8) & 1397(8) & 12 & 0.25\\
B33 & 57645.447641274353 & 235.247 & 4.0(3) & 1446(7) & 1336(7) & 8 & 0.17\\
B34 & 57645.448801855993 & 100.274 & 1.47(6) & 1569(7) & 1373(7) & 18 & 0.14\\
B35 & 57645.449986061139 & 102.315 & 9.2(5) & 1543(10) & 1320(10) & 12 & 0.24\\
B36 & 57645.449986366482 & 0.026 & 2.4(2) & 1730 & 1579(9) & 9 & 0.12\\
B37 & 57645.453425201558 & 297.115 & 2.8(1) & 1506(6) & 1364(6) & 13 & 0.20\\
B38 & 57645.453638065388 & 18.391 & 6.2(3) & 1730 & 1611(5) & 14 & 0.3\\
B39 & 57645.462105667626 & 731.6 & 7.0(3) & 1449(5) & 1315(5) & 15 & 0.27\\
B40 & 57645.464187553633 & 179.875 & 3.7(3) & 1730 & 1504(20) & 10 & 0.09\\
B41 & 57645.474447230416 & 886.436 & 4.8(4) & 1730 & 1600(9) & 12 & 0.17\\
\bottomrule

\end{longtable}
\bibliographystyle{yahapj}
\bibliography{references}

\end{document}